\documentclass[useAMS,referee]{biom}
\def\bSig\mathbf{\Sigma}

\usepackage{amsmath}
\usepackage{titlesec}
\usepackage{graphicx}
\usepackage[colorlinks=true,urlcolor=blue,citecolor=purple,linkcolor=blue,bookmarks=true]{hyperref}
\usepackage{xurl}

\title[Self-adapting Mixture Prior]{SAM: Self-adapting Mixture Prior to Dynamically Borrow Information from Historical Data in Clinical Trials}

\author{Peng Yang$^{1, 4}$, Yuansong Zhao$^{2}$, Lei Nie$^{3}$, Jonathon Vallejo$^{3}$, Ying Yuan$^{4,*}$\email{yyuan@mdanderson.org},  \\
$^{1}$Department of Statistics, Rice University, Houston, TX, U.S.A. \\
$^{2}$Department of Biostatistics, The University of Texas Health Science Center, Houston, TX, U.S.A. \\
$^{3}$Center for Drug Evaluation and Research, Food and Drug Administration (FDA), Silver Spring, MD, U.S.A.\\
$^{4}$Department of Biostatistics, The University of Texas MD Anderson Cancer Center, Houston, TX, U.S.A. 
}

\begin{document}









\label{firstpage}


\begin{abstract}
Mixture priors provide an intuitive way to incorporate historical data while accounting for potential prior-data conflict by combining an informative prior with a non-informative prior.  However, pre-specifying the mixing weight for each component remains a crucial challenge. Ideally, the mixing weight should reflect the degree of prior-data conflict, which is often unknown beforehand, posing a significant obstacle to the application and acceptance of mixture priors. To address this challenge, we introduce self-adapting mixture (SAM) priors that determine the mixing weight using likelihood ratio test statistics or Bayes factors. SAM priors are data-driven and self-adapting, favoring the informative (non-informative) prior component when there is little (substantial) evidence of prior-data conflict. Consequently, SAM priors achieve dynamic information borrowing. We demonstrate that SAM priors exhibit desirable properties in both finite and large samples and achieve information-borrowing consistency. Moreover, SAM priors are easy to compute, data-driven, and calibration-free, mitigating the risk of data dredging. Numerical studies show that SAM priors outperform existing methods in adopting prior-data conflicts effectively. We developed R package ``SAMprior" and web application that are freely available at CRAN and \url{www.trialdesign.org} to facilitate the use of SAM priors.
\end{abstract}



\maketitle


%

\section{Introduction}
\label{s:intro}
Leveraging historical or real-world data has tremendous potential to enhance the efficacy and practicability of clinical trials, especially in the context of rare diseases, pediatric trials involving extrapolation from adult to pediatric populations, and bridging studies that extend findings from one region or ethnic group to another (ICH, 2022). The 21st Century Cures Act, enacted in 2016, recognizes the value of real-world evidence in supporting the approval of new indications for approved drugs and fulfilling post-approval study requirements. To that end, the FDA has issued guidelines on the use of real-world evidence in regulatory decision-making for medical devices (FDA, 2017), as well as a draft guidance on submitting documents using real-world data and evidence to the FDA for drugs and biologics in the industry (FDA, 2019).

One of the most intuitive ways to incorporate historical data into a new trial is to use an informative prior constructed based on the historical data.
However, this method can result in bias and inflated type I errors if the current trial data conflict with the prior. To address this issue, mixture priors provide an intuitive approach to acknowledge the possibility of prior-data conflict and enhance the robustness of information borrowing. In its simplest form, a mixture prior mixes an informative prior with a non-informative or vague prior, assigning a certain mixing weight to each component. The informative prior corresponds to full information borrowing, while the non-informative prior corresponds to little-to-no information borrowing. An example of a prominent mixture prior is the robust meta-analytic predictive (MAP) prior proposed in the seminal work by Schmidli et al. (2014), which mixes a MAP prior with a vague prior.

Determining the mixing weight in the mixture prior is critically challenging. The ideal weight should reflect the degree of relevance of the historical data to the new trial, or the congruence between the two data sets. Unfortunately, this information is typically unknown at the outset of the study, making it difficult to pre-specify the weight in the study protocol. If the weight is overly aggressive towards the informative prior component, excessive information borrowing may occur, leading to substantial bias when there is prior-data conflict. Conversely, if the weight is overly conservative towards the non-informative prior component, the borrowing of historical data may be limited, undermining the purpose of incorporating the historical data. This issue has been a significant barrier to the adoption of mixture priors by investigators and regulatory agencies alike.

In this paper, we propose a solution to the aforementioned barrier by introducing self-adapting mixture (SAM) priors. Our approach addresses the limitation of fixed-weight mixture priors by utilizing a data-driven and self-adapting mixture weight. The SAM prior dynamically favors the informative (non-informative) prior component when there is little (substantial) evidence of prior-data conflict. 
Notably, the procedure of constructing the SAM prior can be fully pre-specified in the study protocol. We show that SAM priors possess desirable finite-sample and large-sample properties, ensuring information-borrowing consistency, and outperforming existing fixed-weight mixture priors.

In addition to mixture priors, several other approaches have been proposed to account for the possibility of prior-data conflict in information borrowing.
Ibrahim and Chen (2000, 2003) proposed power priors, which discount the historical data using a power parameter to acknowledge the possibility of prior-data conflict.  Hobbs et al. (2011) proposed commensurate priors that control information borrowing based on the commensurability between historical data and current data, while Hobbs et al. (2012, 2013) extended this approach to generalized linear models and utilized it to adjust the randomization ratio adaptively in randomized controlled trials.  Recently, Jiang, Nie, and Yuan (2022) proposed elastic priors that proactively control the amount of information borrowing based on the similarity between historical and current data through an elastic function. 

In addition to these prior-based approaches, Bayesian hierarchical models (BHMs) provide a flexible framework for borrowing information among multiple parallel subgroups or data resources. Thall et al. (2003) and Berry et al. (2013) proposed using BHMs to borrow information from different subgroups. Neuenschwander et al. (2016) used a mixture of exchangeable and non-exchangeable priors to improve the robustness of BHMs. Chu and Yuan (2018) proposed a calibrated BHM to enhance the dynamic borrowing of BHMs. Kaizer et al. (2018) developed a multisource exchangeability BHM to accommodate heterogeneity across multiple data sources. Jiang et al. (2021) proposed an easy-to-implement clustered BHM that clusters multiple arms  before borrowing information within each cluster using a BHM.

The remainder of this article is organized as follows. In Section 2, we propose SAM priors and study their statistical properties. In Section 3, we evaluate the operating characteristics of the proposed method using simulation and provide an application example in Section 4. We conclude with a brief discussion in Section 5.

\vspace{-0.5cm}
\section{Methods}
\subsection{Mixture prior}
Consider a new randomized controlled trial (RCT) comparing a new test treatment with a control, where relevant historical data are available only to the control arm. The objective is to incorporate the historical data into the analysis of the new trial and obtain posterior inference for the parameter of interest $\theta$ representing the treatment effect. Let $y$ denote the endpoint of interest,  $D_h=(y_{h1}, \cdots, y_{h n_h})$ denote the historical data collected from $n_h$ independent subjects, and $D=(y_1, \cdots, y_n)$ denote the  new trial data collected from $n$ independent subjects in the control arm. As relevant historical data are only available to the control, we will focus on the posterior inference of $\theta$ for the control arm.  The posterior inference for the treatment arm will be done using standard Bayesian methods, e.g., using a conventional non-informative or vague prior. To simplify the presentation, following Schmidli, et al. (2014), we assume that no nuisance parameters are present. Although we focus on RCTs with information borrowing for the control, the proposed method is also applicable to RCTs aiming to borrow information for both treatment and control arms, such as pediatric trials with relevant adult data or bridging trials with historical data from other regions or ethnic groups, as well as single-arm trials with relevant historical data. 

We assume that an informative prior, denoted by $\pi(\theta|D_h)$, has been constructed based on $D_h$ using a certain methodology. For example, when $D_h$ is from a single study, $\pi(\theta|D_h)$ can be constructed by applying Bayes' rule, i.e., $\pi(\theta|D_h)  \propto \pi_0(\theta)p(D_h | \theta)$, where $\pi_0(\theta)$ denotes a non-informative or vague prior. When $D_h$ consists of multiple studies, a reasonable choice for $\pi(\theta|D_h)$ is the MAP prior. Spiegelhalter et al. (2004) and Falconer et al. (2022) discussed various ways to construct an informative prior based on historical data. Importantly, the proposed methodology is not tied to any particular informative prior construction method. For notational brevity, we shorthand $\pi(\theta|D_h)$ as $\pi_1(\theta)$. 

For $\pi_0(\theta)$, Schmidli et al. (2014) recommended Jeffreys' or the uniform prior for binary endpoints and unit information priors for other endpoints (Kass and Wasserman, 1995). The unit information prior is a vague prior that contains information equivalent to that of one observation. Kass and Wasserman (1995) provide a formal definition and rationale for using the unit information prior as a reference or automatic prior.

To acknowledge the possibility of prior-data conflict and improve the robustness of the inference, Schmidli et al. (2014) proposed fixed-weight mixture priors: 
\begin{equation}
\pi_m(\theta) = \tilde{w} \pi_1(\theta) + (1-\tilde{w}) \pi_0(\theta), \label{fixmixture}
\end{equation}
where $\tilde{w}$ is a prespecified {\it fixed} mixing weight, representing the prior probability of no prior-data conflict between $D_h$ and $D$, which controls the degree of information borrowing from $D_h$. When $\tilde{w}=1$, $\pi_m(\theta)$ achieves full information borrowing; and when $\tilde{w}=0$, $\pi_m(\theta)$ does not borrow any information from $D_h$.  

Determining the appropriate value of $ \tilde{w}$ is critically challenging. Ideally, $ \tilde{w}$ should truly reflect the level of prior-data conflict or the degree to which $D_h$ is congruent with $D$. However, this information is often unknown in advance. Setting $ \tilde{w}$ to a very high value can lead to excessive information borrowing, resulting in substantial bias that may cause inflation of type I errors or loss of power. Conversely, setting $ \tilde{w}$ too low may limit the amount of information borrowing, reducing the potential power gain and rendering the inclusion of $D_h$ pointless. As $\pi_1(\theta)$ is constructed to encapsulate the information contained in $D_h$, the terms ``prior-data conflict" and ``incongruence between $D_h$ and $D$" will be used interchangeably in the following discussion.

Because $\pi_1(\theta)$ and $\pi_0(\theta)$ are independent, the posterior resulting from the fixed-weight mixture prior (\ref{fixmixture}) is a mixture of two posteriors, with weights updated by normalizing constants (Bernardo and Smith, 1994). It is important to recognize that this conjugate structure should not be interpreted as the fixed-weight mixture prior to dynamically adjusting its weight based on the prior-data conflict. In accordance with Bayes' rule, the information encompassed within the fixed-weight mixture prior will be fully incorporated into the posterior at its face value regardless of the degree of prior-data conflict, akin to any other informative priors. For instance, if the fixed-weight mixture prior contains the information equivalent to 100 patients, then that information will be fully incorporated into the posterior inference. The true benefit of mixture priors lies in their ability to offer heavy-tailed distributions, making them less sensitive to prior-data conflict. Schmidli et al. (2014) aptly employed the term ``robust" instead of ``dynamic", accurately capturing the essence of this method. 

\subsection{Self-adapting mixture prior}
We introduce a data-driven approach for determining the value of $\tilde{w}$, which yields SAM priors capable of dynamically adjusting the mixture weight based on the extent of the prior-data conflict. Let $\theta_h$ denote the treatment effect associated with $D_h$, which may be identical to or substantially different from $\theta$. We define the {\it clinically significant difference} (CSD) in the treatment effect as $\delta$, such that if $|\theta_h-\theta| \ge \delta$, then $\theta_h$ is regarded as clinically distinct from $\theta$, and it is therefore inappropriate to borrow any information from $D_h$. The appropriate value of $\delta$ should be determined through consultation with domain experts and regulatory bodies, and may vary depending on the disease or condition under study.  It is important to note that CSD should not be conflated with the minimal clinically significant difference, which represents the smallest improvement that is clinically meaningful. As CSD represents the threshold beyond which no information should be borrowed, it is typically greater than the minimal clinically significant difference. For the sake of brevity, we assume that CSD is the same in both directions (i.e., $\theta >\theta_h$ or $\theta<\theta_h$), although the proposed method can be readily adapted for scenarios in which CSD varies in these directions. The inclusion of clinical expertise and knowledge to guide and regulate the behavior of information-borrowing is a key attribute and advantage of this methodology. A similar approach was previously employed in the elastic prior (Jiang et al., 2023). Therefore, we refer to the SAM prior and elastic prior as supervised information-borrowing methods, while power prior, commensurate prior and mixture prior as unsupervised information-borrowing methods.

To begin, we introduce two hypotheses, represented by $H_0$ and $H_1$, as follows:
\begin{equation} \label{hypotheses}
H_0: \theta = \theta_h, \qquad H_1: \theta = \theta_h+\delta \,\, \mathrm{or} \,\, \theta = \theta_h-\delta.
\end{equation}
We temporarily assume that $\theta_h$ is known. Under $H_0$, $\pi_1(\theta)$ and $D$ are consistent and exhibit no prior-data conflict, thus it is appropriate to employ $\pi_1(\theta)$ to borrow information from $D_h$. Conversely, under $H_1$, the treatment effect of $D$ and $D_h$ differ to such a degree that no information should be borrowed, and the posterior inference of $\theta$ should instead utilize $\pi_0(\theta)$. Given $D$ and $\theta_h$, the extent of information borrowing can be determined by the relative likelihood of $H_0$ and $H_1$ being accurate, which can be quantified using the likelihood ratio test (LRT) statistic:
 \begin{equation} \label{LRT}
 R  =   \frac{p(D | H_0, \theta)}{p(D | H_1, \theta)}  = \frac{p(D | \theta=\theta_h)}{{\rm max}\{p(D |\theta= \theta_h+\delta), p(D | \theta=\theta_h-\delta)\}},
 \end{equation}
where $p(D | \cdot )$ denotes the likelihood. In the denominator of equation (\ref{LRT}),  we opt to use the maximum of $p(D | H_1, \theta)$  for less aggressive information borrowing to ensure better control of bias and minimize type I error inflation. An alternative Bayesian choice to measure the relative likelihood of $H_0$ and $H_1$ being accurate is the posterior probability ratio (PPR):
\begin{equation}
R = \frac{p(H_0 | D)}{p(H_1 | D)} = \frac{p(H_0)}{p(H_1)} BF, 
\end{equation}
where $p(H_0)$ and $p(H_1)$ are the prior probabilities of $H_0$ and $H_1$ being true, respectively, which are equivalent to $\tilde{w}$ and $1-\tilde{w}$ in the fixed-weight mixture prior (\ref{fixmixture}). $BF$ is the Bayes factor that in this case is the same as the LRT in (\ref{LRT}). LRT is a fully data-driven approach and is preferred when investigators aim to avoid the subjectivity of prior specification on prior-data conflict, as required by the fixed-weight mixture priors. PPR is a partially data-driven method that is useful when investigators want to incorporate prior information on the prior-data conflict, while also desiring data-dependent correction when the prior is misspecified. When a non-informative prior $p(H_0)=p(H_1)$ is used for PPR, the two approaches are equivalent.  

The SAM prior is then defined as 
\begin{equation}
\pi_{sam}(\theta) = w \pi_1(\theta)  +  (1-w) \pi_0(\theta), \label{sam}
\end{equation}
where 
\begin{equation}
w \propto R =  \frac{R}{1 + R}. \label{adptweight}
\end{equation}
As the level of prior-data conflict increases, $R$ decreases,  resulting in a decrease in the weight $w$ assigned to information borrowing. Thus, $\pi_{sam}(\theta)$ has the ability to self-adjust based on the degree of prior-data conflict.   In practice, $\theta_h$ is unknown and can be substituted with an estimate, such as the posterior mean estimate $\hat{\theta}_h=E(\pi_1(\theta))$ or the maximum likelihood estimate, to calculate $w$. 

One significant advantage of the SAM prior, as a mixture prior, is that they are often analytically tractable. This substantially simplifies posterior inference, as demonstrated in the subsequent section. The computational requirements of utilizing SAM priors are similar to those of fixed-weight mixture priors. As Schmidli et al. (2014) have emphasized, the tractability of the analysis lowers the barrier to implementation and enables the rapid evaluation of operating characteristics. To facilitate the use of the SAM priors, we have developed both R package ``SAMprior" and a web application, which are available for free on CRAN and \url{www.trialdesign.org}, respectively.

The SAM prior is an empirical Bayes method as $w$ depends on $D$. Additionally, we investigate three alternative approaches: (a) a fully Bayesian approach by assigning $w$ a uniform prior, and two modifications of the fixed-weight mixture prior by assigning $\pi_0(\theta)$ a bimodal prior including (b) the inverse moment (iMOM) prior (Johnson and Rossell, 2010) and (c) a mixture of two beta/normal priors with modes at $\theta \pm \delta$. Methods (b) and (c) aim to account for $H_1$ and CSD in the fixed-weight mixture prior. Despite their apparent appeal, none of these approaches perform well (see Supporting Information). 
Similar challenges have been observed for power prior and commensurate prior (Neuenschwander et al., 2000; Jiang et al., 2023).

\vspace{-12pt}
\subsubsection{Examples with binary and normal endpoints}
Consider a binary endpoint $y \sim Bernoulli (\theta)$, where $\theta$ is the response rate. Let $x =\sum_{i}^n y_i$ denote the number of responses among $n$ subjects treated in the control arm. Given a single historical dataset $D_h$ with $x_h$ responses out of $n_h$ subjects, a commonly used informative prior is 
$
\pi_1(\theta) = Beta(a+x_h, b+n_h-x_h), \label{betaprior}
$
constructed by applying a beta-binomial model to $D_h$ with a vague/non-informative prior $\pi_0(\theta) = Beta(a, b)$. Schmidli et al. (2014) recommended the uniform prior by setting $a=b=1$.

Let $\hat{\theta}_h = (a+x_h)/ (a+b+n_h)$ denote the estimate of $\theta_h$ implied by $\pi_1(\theta)$. The SAM prior is given by
$
\pi_{sam}(\theta) = w Beta(a+x_h, b+n_h-x_h) + (1-w) Beta(a, b),
$
where $w=R/(1+R)$ with 
$R = \frac{{\hat{\theta}_h}^x (1-\hat{\theta}_h)^{n-x}}{ {\rm max} \left\{ (\hat{\theta}_h+\delta)^x (1-\hat{\theta}_h-\delta)^{n-x}, (\hat{\theta}_h- \delta)^x (1-\hat{\theta}_h+\delta)^{n-x}\right\}     } .
$
Owing to its conjugacy, given $\pi_{sam}(\theta)$ and the trial data $D$, the posterior of $\theta$ is given by
$p(\theta | D, D_h) = w^* Beta(a+x_h+x, b+n_h+n-x_h-x) + (1- w^*) Beta(a+x, b+n-x),$
where $w^*$ is given by $w^*  = \frac{w z_1}{w z_1 + (1-w) z_0}$, $z_0  = \frac{B(a + x, n - x + b)}{B(a,b)}$, and $z_1  = \frac{B(a + x_h + x, b + n_h + n - x_h - x)}{B(a + x_h, b + n_h - x_h)}$, with $B(\cdot, \cdot)$ standing for beta function. In the case that $D_h$ consists of multiple potentially heterogeneous datasets, the MAP prior $\pi_{_{MAP}}(\theta)$ can be used as $\pi_1(\theta)$. The resulting SAM prior is given by
$\pi_{sam}(\theta) = w \pi_{_{MAP}}(\theta) + (1-w) Beta(a, b)$, where $w$ is the same as above with  $\hat{\theta}_h = \int \theta \pi_{_{MAP}}(\theta) d \theta$.

We briefly discuss the SAM prior for a continuous endpoint $y \sim N(\theta, \sigma^2)$. Let $\bar{y}_h$ and $s$ denote the sample mean and standard error of $D_h$. We take $\pi_1(\theta) = N(\bar{y}_h, s^2/n)$, obtained as $p(\theta|D_h)$ with non-informative prior $p(\theta, \sigma^2) \propto (\sigma^2)^{-1}$. Strictly speaking,  $p(\theta|D_h)$ follows a $t$ distribution with the degree of freedom of $n_h-1$, but it can be well approximated by a normal distribution given that $n_h$ is typically moderate and large. We use the unit information prior as $\pi_0(\theta)$, i.e., $\pi_0(\theta) \sim N(\bar{y}_h, \hat{\sigma}^2)$, where $\hat{\sigma}^2=s^2$ is an estimate of $\sigma^2$. Here, the non-informative prior used to obtain $\pi_1(\theta)$ is different from $\pi_0(\theta)$, exemplifying the flexibility of the SAM prior, but $\pi_1(\theta)$ can be built upon $\pi_0(\theta)$. 
The SAM prior is given by
$
\pi_{sam}(\theta) = w N(\bar{y}_h, s^2/n) + (1-w) N(\bar{y}_h, s^2),
$
with $w=R/(1+R)$ and 
$R = \exp \left\{ -\frac{1}{2} \sum_{i=1}^n\left( \frac{y_i-\hat{\theta}_h}{\hat{\sigma}}\right)^2 - {\rm max}\left\{-\frac{1}{2}\sum_{i=1}^n\left( \frac{y_i-\hat{\theta}_h-\delta}{\hat{\sigma}}\right)^2, -\frac{1}{2}\sum_{i=1}^n\left( \frac{y_i-\hat{\theta}_h+\delta}{\hat{\sigma}}\right)^2 \right\}
\right\},
$
where $\hat{\theta}_h =\bar{y}_h$ and $\hat{\sigma}$ is the sample variance estimate based on $D$ or the pooled sample estimate based on $D$ and $D_h$.

The SAM prior method can also be extended to the analysis of survival endpoints. More details can be found in the Supporting Information. 

\vspace{-1cm}
\subsubsection{Statistical properties}
 SAM prior has the following desirable large-sample properties. The proof is provided in the Web Appendices. 

\noindent{\bf Theorem 1} $\quad$  { The SAM prior (\ref{sam}) converges to $\pi_1(\theta)$ if $D_h$ and $D_c$ are congruent (i.e., $\theta_h=\theta$), and converges to $\pi_0(\theta)$ if $D_h$ and $D_c$ are incongruent (i.e., $|\theta-\theta_h| = \delta$).

\noindent{\bf Corollary 1} $\quad$  The SAM prior is information-borrowing consistent in the sense that when the sample size of $D$ and $D_h$ are large, it achieves full information borrowing contained in $\pi_1(\theta)$ if $D_h$ and $D$ are congruent (i.e., $\theta_h=\theta$), and it discards $D_h$ if $D_h$ and $D$ are incongruent (i.e., $|\theta-\theta_h| = \delta$).

In contrast, some existing information borrowing methods, such as power priors and commensurate priors, may not possess these desirable properties. This is because the observation unit used to estimate the information-borrowing parameter (e.g., power parameter, shrinkage parameter) in these approaches is the dataset rather than the subject (Neuenschwander et al., 2000; Jiang et al., 2023). Increasing the number of subjects does not guarantee the convergence of the estimate of the information-borrowing parameters. Our simulation study, described later, shows that compared to power priors and commensurate priors, SAM priors are more responsive and adaptive to prior-data conflict and exhibit better performance in controlling bias and type I errors in the presence of such conflict. Furthermore, compared to commensurate priors, posterior inference for SAM priors is often simpler and analytically trackable. SAM priors are also more flexible and can seamlessly handle single or multiple $D_h$ by choosing different forms of $\pi_1(\theta)$, as described previously. This flexibility can be challenging for commensurate priors and power priors.
 
Compared to elastic priors, which are also information-borrowing consistent, SAM priors are simpler. Elastic priors require simulation to calibrate the elastic function to achieve desirable finite-sample operating characteristics. In contrast, SAM priors are essentially calibration-free. Once the CSD is elicited, 
SAM priors are fully specified.  

Compared to fixed-weight mixture priors, such as robust MAP priors, SAM priors offer the advantage of adaptivity and self-adjustment of the mixing weight based on the degree of prior-data conflict. This leads to more adaptive information borrowing, resulting in generally smaller bias, mean square errors, and better type I error control in the presence of prior-data conflict, as demonstrated through simulation (see Section \ref{sec.simu}). Additionally, SAM priors are data-driven and remove the need to prespecify mixing weights, thus avoiding selection bias and potential data dredging inherent in fixed-weight mixture priors. This property of SAM priors aligns with the considerations in the draft Guidance for Industry on Interacting with the FDA on Complex Innovative Trial Designs for Drugs and Biological Products, which states: ``Bayesian CID proposals should include a robust discussion of the prior distribution...a Bayesian proposal should also include a discussion explaining the steps the sponsor took to ensure information was not selectively obtained or used. In cases where downweighting or other non-data-driven features are incorporated in a prior distribution, the proposal should include a rationale for the use and magnitude of these features." 

\vspace{-14pt}
\section{Simulation studies} \label{sec.simu}
\subsection{Simulation setting}\label{simuset}
We investigated the operating characteristics of the SAM prior via simulation. We assume that $D_h$ is a single historical dataset, but the results are applicable to $D_h$ that consists of multiple historical datasets. This is because given $\pi_1(\theta)$ and $\hat{\theta}_h$, no matter whether they are obtained based on a single dataset or multiple datasets (e.g., using the MAP prior),  the same SAM prior and thus the same results will be obtained.  We considered both binary and normal endpoints. As shown in Table \ref{tab:binary}, for the binary endpoint we considered three response rates for $D_h$, i.e., $\theta_h=0.4, 0.3$, and 0.2, and generated $D_h$ from $Bernoulli(\theta_h)$ with sample size $n_h = 300, 300$, and 250, respectively. We generated control arm data $D$ from $Bernoulli(\theta)$ and varied the value of $\theta$ to simulate different degrees of prior-data conflict. We generated treatment arm data $D_t$ from $Bernoulli(\theta_t)$ and varied the value of $\theta_t$ to simulate different treatment effect sizes (see Table 1), and assumed 2:1 randomization between the treatment and control arms. When $\theta_h=0.4, 0.3$, and 0.2, we set the control arm sample size $n = 150, 150$, and 125, respectively, and the treatment arm sample size $n_t = 300, 300$, and 250, respectively. The sample sizes were chosen such that the power of the methods under comparison is mostly in the range of 70\% to 90\%. In all simulation scenarios, CSD $\delta=0.1$ and noninformative prior $\pi_0(\theta) \sim Unif(0,1)$ were used.

For the normal endpoint, we generated $D_h$ from $N (\theta_h=0,  \sigma^2)$, the control arm data $D$ from $N (\theta, \sigma^2)$, and the treatment arm data $D_t \sim N(\theta_t, \sigma^2)$, where $\sigma$ is the standard deviation and set as $\sigma=3$. As shown in Table \ref{tab:normal}, we varied the value of $\theta$ to simulate different degrees of prior-data conflict, and varied the value of $\theta_t$ to obtain the treatment effect size $d=(\theta_t-\theta)/\sigma = 0.2$ (small), 0.5 (medium), and 0.8 (large). We assumed 2:1 randomization between the treatment and control arms. Given the small, medium, and large effect sizes, we set $n_h= 350, 60$, and 30;  $n$ = 175, 30, and 12; and $n_t = 350, 60$, and 24. The sample size was chosen such that power of the designs under comparison was mostly between 70\% to 90\%. We set CSD $\delta = 0.2\sigma$, $0.5\sigma$, and $0.8\sigma$ for small, medium, and large effect size setting, respectively, and used a unit information prior (Kass and Wasserman, 1995) as $\pi_0(\theta)$, i.e., $\pi_0(\theta) = N(0, 3)$.
In both binary and normal endpoint simulations, we considered $D_h$ fixed to imitate the practice (e.g., generated $D_h$ once under each setting with the constraint that  $\bar{y}_h = \theta_h$) and generated the replicates of the trial data $D$ and $D_t$. Under each simulation configuration, 2000 simulations were conducted. 

We compared the SAM prior, constructed using LRT, to the following methods: (1) a conventional non-informative prior (NP) approach that ignores the historical data and generates the posterior based on $\pi_0(\theta)$, (2) a fixed-weight mixture prior with $\tilde{w}=0.5$ (Mix50), (3) a power prior (PP) with a uniform prior $Unif(0, 1)$ for the power parameter, and (4) a commensurate prior (CP) with $log(\tau) \sim Unif(-30, 30)$ where $\tau$ is the shrinkage parameter (Hobbs, et al., 2011). The same criterion is used across the methods to evaluate the efficacy of the treatment: the treatment is deemed superior to the control if $\Pr(\theta_t-\theta> 0 |D, D_t, D_h)>C$, where the probability cutoff $C$ is calibrated independently for each method using simulation such that the type I error is 5\% when the null (i.e., $\theta_t=\theta$) is true and that $D_h$ and $D$ are congruent (i.e., $\theta_h=\theta$). The values of $C$ are provided in the Supporting Information. 

\vspace{-0.4cm}
\subsection{Simulation results}
Figure \ref{fig:binary} depicts the relative bias and relative mean square error (RMSE) of the posterior mean estimate of $\theta$ under SAM, Mix50, PP, and CP for the binary endpoint, with NP serving as the reference. The relative bias is the difference between the bias of a method and the bias of NP, while the RMSE is the difference between the mean square error (MSE) of a method and the MSE of NP. Figure \ref{fig:binary}A indicates that SAM exhibits a uniformly smaller bias than Mix50, CP, and PP across the range of $\theta$, implying its better adaptation to prior-data conflict than the other methods. SAM's bias diminishes more rapidly than the other methods as the prior-data conflict grows. SAM's superiority arises from its capability to self-adjust $w$ based on the extent of prior-data conflict, as shown in Figure \ref{fig:weight}. The value of $w$ is highest when $\theta=\theta_h$, but quickly decays as $\theta$ moves away from $\theta_h$, reducing information borrowing and therefore bias. Figure \ref{fig:binary}B demonstrates a comparable pattern in RMSE. When $\theta$ is equal to or near $\theta_h$, all methods display similar reductions in MSE due to information borrowing. When $\theta$ is very near to $\theta_h$, SAM's MSE is slightly higher than other methods because SAM is data-driven and leads to slightly less borrowing after accounting for data uncertainty. When $\theta$ deviates from $\theta_h$, SAM's RMSE is substantially lower than the other methods.

Table \ref{tab:binary} summarizes the type I error rate and power of the different methods. All methods are calibrated to control the type I error rate at 5\% under the null hypothesis where $\theta_t=\theta$ (scenario 1.1). In scenarios 1.2-1.4 where the treatment is effective, SAM exhibits substantial power gain over NP and shows higher or comparable power to Mix50, PP, and CP. For example, in scenario 1.2, the power of SAM is 22.6\% higher than that of NP. Scenarios 1.5-1.8 represent the situations where there is prior-data conflict between $D_h$ and $D$. In scenarios 1.5-1.6 where the treatment is not effective, SAM outperforms Mix50, PP, and CP in controlling type I errors. For instance, in scenario 1.6, the type I error of SAM is 8.4\%, whereas the type I error of Mix50 is 12.2\%, and the type I errors of PP and CP are 26.2\% and 20.0\%, respectively. In scenarios 1.7-1.8 where the treatment is effective, SAM exhibits higher power to detect the treatment effect than that of PP and CP. For example, in scenario 1.8, the power of SAM is 73.9\%, while the power of Mix50, PP, and CP are 60.0\%, 44.6\%, and 47.4\%, respectively. The results under $\theta_h=0.3$ (scenarios 2.1-2.8) and $\theta_h=0.2$ (scenarios 3.1-3.8) are similar to scenarios 1.1-1.8.

Figure \ref{fig:normal} displays the relative bias and relative mean square error (RMSE) for a normal endpoint, while Table \ref{tab:normal} summarizes the corresponding type I error and power. The findings are largely consistent with those observed for the binary endpoint. Specifically, SAM demonstrates superior adaptability to prior-data conflict, as evidenced by its uniformly lower bias than Mix50, CP, and PP. When little to no prior-data conflict is present, SAM yields a similar reduction in MSE as the other methods, but it produces lower MSE in scenarios with substantial prior-data conflict. Furthermore, SAM yields comparable power gain to Mix50, CP, and PP when there is little to no prior-data conflict, while demonstrating better type I error control and higher power in the presence of prior-data conflict.

We also investigated the operating characteristics of the SAM prior for a survival endpoint. The results are generally consistent with those of the normal and binary endpoints, see the Supporting Information for details.

\vspace{-0.4cm}
\subsection{Sensitivity analysis}
To assess the sensitivity of the SAM prior to the specification of CSD, we considered $\delta=0.15$ for the binary endpoint. The results are similar to those with $\delta=0.1$ and provided in the Supporting Information (Figure S1 and Table S1). For the normal endpoint, the primary simulation considered varying CSDs, showing similar robustness. Additionally, we examined how the mixture weight varied in relation to the CSD for both binary and continuous endpoints and presented the findings in the Supporting Information (Figure S4).  The results demonstrate that the mixture weights are peaked within the CSD and drop significantly beyond that as the prior-data conflict increases. We also evaluated the performance of the SAM prior constructed using PPR. The results are similar to these described above and provided in the Support Information.

\section{Application}
Ankylosing spondylitis is a chronic immune-mediated inflammatory disease characterized by spinal inflammation, progressive spinal rigidity, and peripheral arthritis. Consider a randomized clinical trial to compare a treatment with a control in patients with ankylosing spondylitis. The primary efficacy endpoint is binary, indicating whether a patient achieves $\ge$ 20\% improvement at week six according to the Assessment of SpondyloArthritis International Society criteria (Anderson, et al., 2001). Nine historical datasets are available for the control; see the Supporting Information (Table S5) for the dataset (Wang, et al., 2021). The response rate of the historical controls varies from 0.17 to 0.47, with the sample size ranging from 6 to 153. The MAP prior was constructed and served as $\pi_1(\theta)$ to incorporate the historical control data. The resulting MAP prior is approximated by a mixture of conjugate priors, given by $\pi_1(\theta) = 0.63 Beta (42.5, 77.2) + 0.37 Beta (7.2, 12.4)$. The mean of MAP prior is $\hat{\theta}_h = 0.36$. 

To evaluate the performance of the SAM prior for this trial, we conducted simulations by generating the control arm data $D$ from $Bernoulli(\theta)$ while varying the value of $\theta$ to simulate scenarios with varying degrees of prior-data conflict. The treatment arm data $D_t$ were generated from $Bernoulli(\theta_t)$ with different values of $\theta_t$ representing different treatment effect sizes. The total sample size of the trial is 105, randomized in a 1:2 ratio to the control and treatment arms. We considered the treatment arm to be superior to the control arm if $\Pr(\theta_t-\theta> 0 |D, D_t, D_h)>C$, where $C$ was calibrated by simulation to control the type I error at 5\% under the null hypothesis. We set $\delta = 0.2$ and $\pi_0(\theta) \sim Unif (0,1)$. We compared the performance of the SAM prior with the robust MAP prior. For the robust MAP prior, we considered two choices of weight, namely $\tilde{w} = 0.5$ or 0.9, denoted as Mix50 and Mix90, respectively. We also used the NP approach that ignores the historical data as a reference.

Figure \ref{fig:app} displays the relative bias and MSE for Mix50 and SAM based on 2000 simulations. SAM has a uniformly lower bias than Mix50. Although the RMSE of SAM is slightly higher than Mix50 when prior-data conflict is minor, it is substantially lower than Mix50 when prior-data conflict presents. Table \ref{tab:app} summarizes the type I error rate and power of the two methods. When prior-data conflict is minimal, SAM yields comparable power to the corresponding robust MAP, which is substantially higher than NP.  However, when prior-data conflict is present, SAM demonstrates better type I error control and higher power than the robust MAP. For example, in scenario 6 the type I error of SAM is 10.3\%, whereas the type I error of Mix50 and Mix90 is 12.8\% and 25.0\%, respectively. In scenario 8, the power of SAM is 11.3\% and 28.7\% higher than that of Mix50 and Mix90. The superior performance of SAM can be attributed to its adaptive self-adjustment of the mixing weight according to prior-data conflict (see Figure \ref{fig:app}C). We noted that one trial (Baeten 2013) used to generate the MAP prior has a small sample size ($n=6$). We conducted a sensitivity analysis by excluding that dataset. The results are similar and reported in the Supporting Information.

\section{Conclusion}
We have proposed SAM priors as a means of achieving dynamic information borrowing from historical data. 
The SAM prior is both data-driven and self-adapting, favoring the informative (non-informative) prior component when there is little (substantial) evidence of prior-data conflict. This approach helps to circumvent selection bias and potential data dredging, which may compromise the fixed-weight mixture prior approach. Our findings demonstrate that SAM priors possess desirable finite-sample and large-sample properties, ensuring information-borrowing consistency. Simulation shows that SAM outperforms fixed-weight mixture priors and other existing methods, demonstrating better adaptation to prior-data conflict.

The SAM prior is highly flexible, enabling researchers to utilize different existing methods to construct the informative prior component $\pi_1(\theta)$. This customization facilitates targeted information borrowing. For instance, in a small pediatric trial that seeks to leverage information from a large adult dataset containing thousands of observations, it may be preferable to limit the maximum amount of information borrowed from historical data. To achieve this, researchers can either inflate the variance of $p(\theta|D_h)$ or employ a power prior (to down-weight the likelihood) when constructing $\pi_1(\theta)$.

We view the requirement for CSD specification as a significant advantage of the SAM prior, rather than a drawback, compared to fixed-weight mixture priors and other methods such as power and commensurate priors. Information-borrowing methods inevitably entail trade-offs between type I error inflation and power, as well as between bias and efficiency. Therefore, clinical knowledge and judgment, such as CSD specification, are crucial for informing the analysis and design and regulating the operating characteristics of the method. This reflects clinical practice and accommodates the unique characteristics of individual trials. Furthermore, our sensitivity analysis reveals that SAM priors are not significantly impacted by certain variations in CSD, reinforcing their robustness. 

Some may worry that the SAM prior doubly utilizes the data, like other empirical Bayes approaches. In the context of borrowing information through shrinkage estimates, empirical Bayes has been shown to perform well and often yield more sensible results than full Bayes (Efron and Morris, 1971, 1972; Carlin and Louis, 2000). Our simulation results also support this conclusion.

Measuring the amount of information borrowed from historical data is often of interest, with effective sample size (ESS) being a common metric. However, the calculation of ESS for mixture priors presents some challenges. Wiesenfarth and Calderazzo (2020) studied several approaches and found that the method of Morita et al. (2008) is unsuitable for mixture priors due to the limited characterization of information in mixture distributions. Alternative approaches, such as Schmidli et al. (2014) and Gravestock and Held (2019), better capture the characteristics of mixture distributions but are not data-dependent. Wiesenfarth and Calderazzo (2020) and Reimherr et al. (2021) proposed data-dependent ESS approaches. For SAM priors, we prefer the latter two methods as they align with the objective of achieving dynamic borrowing. However, ESS calculation is a complex concept that depends on the choice of a reference non-informative prior and the measure of information, for which there is no consensus. Different approaches may yield different ESS values. Further research is needed to appropriately measure the information of mixture priors.

The SAM prior does not currently incorporate covariate information, which could further improve the efficiency of information borrowing. Along the line of Wang, et al. (2019), one practical solution is to first use propensity score matching to identify a subset of ${D}_h$ that is comparable to the current trial population based on covariates, and then apply the SAM prior to borrow information from the subset. This propensity score-integrated approach increases the chance of borrowing information by leveraging more congruent data, while also reducing the impact of violating the assumption of no non-measured confounders required by the propensity score method. This approach possesses a double robustness property. However, it is important to choose an appropriate posterior probability cutoff $C$ to control type I error given the added step of propensity score matching. In addition, using patient-level covariates can standardize other elements that may make the data more congruent, such as the length of follow-up for the response and response criteria, which will be the focus of future research.

\vspace*{-16pt}
\section*{ACKNOWLEDGMENTS}
Yang was partially supported by Award Number 5U01DK108328 from National Health Institute. Yuan was partially supported by Award Number P50CA221707, P50CA127001, and CA016672 from the National Cancer Institute, and Bettyann Asche Murray Distinguished Professorship. This article reflects the views of the author, and it should not be construed to represent views or policies of the FDA. \vspace*{-8pt}

\vspace*{-16pt}
\section*{DATA AVAILABILITY STATEMENT}
The data and software that support the findings of this paper are available in the Supporting Information section of this paper.

\vspace*{-16pt}
\section*{SUPPORTING INFORMATION}
Web Appendices, Tables, Figures, and data referenced in Sections 2, 3, and 4 are available with this paper at the Biometrics website on Wiley Online Library. Code for implementing the proposed method is available as the R package ``SAMprior" at CRAN: \url{https://CRAN.R-project.org/package=SAMprior} and GitHub: \url{https://github.com/pengyang0411/SAMprior}.

\clearpage
\begin{table}
\begin{center}
\caption {Type I error rate and power when using a non-informative prior (NP), SAM prior, mixture prior with $\tilde{w}=0.5$ (Mix50), power prior (PP), and commensurate prior (CP) with a binary endpoint.} \label{tab:binary}
\renewcommand\arraystretch{1.12}
\begin{tabular}{ccccccccc}
\hline
\textbf{Scenario} & $\theta$ & $\theta_t$ & \textbf{NP} &  \textbf{SAM} &  \textbf{Mix50} & \textbf{PP} & \textbf{CP} \\ \hline
\multicolumn{8}{c}{\textbf{Case 1: $\theta_h=0.4$}} \\
 \multicolumn{3}{l}{\textbf{Congruent}} &&&&&\\
 1.1* & 0.4 & 0.4 & 0.051 & 0.051 & 0.050 & 0.050 & 0.051 \\
1.2 & 0.4 & 0.5 & 0.636 & 0.862 & 0.878 & 0.875 & 0.874 \\
1.3 & 0.41 & 0.51 & 0.655 & 0.866 & 0.903 & 0.904 & 0.910 \\
1.4 & 0.38 & 0.48 & 0.636 & 0.822 & 0.828 & 0.820 & 0.817 \\
 \multicolumn{3}{l}{\textbf{Incongruent}} &&&&\\
1.5* & 0.5 & 0.5 & 0.056 & 0.160 & 0.221 & 0.271 & 0.329 \\
1.6* & 0.55 & 0.55 & 0.056 & 0.084 & 0.122 & 0.262 & 0.200 \\
1.7 & 0.3 & 0.4 & 0.657 & 0.652 & 0.480 & 0.490 & 0.413 \\
1.8 & 0.25 & 0.35 & 0.690 & 0.739 & 0.600 & 0.446 & 0.474 \\
\hline
\multicolumn{8}{c}{\textbf{Case 2: $\theta_h=0.3$}} \\
\multicolumn{3}{l}{\textbf{Congruent}} &&&&&\\
2.1* & 0.3 & 0.3 & 0.050 & 0.051 & 0.050 & 0.051 & 0.050 \\
2.2 & 0.3 & 0.4 & 0.657 & 0.888 & 0.894 & 0.890 & 0.902 \\
2.3 & 0.31 & 0.41 & 0.649 & 0.882 & 0.908 & 0.912 & 0.912 \\
2.4 & 0.28 & 0.38 & 0.667 & 0.852 & 0.854 & 0.839 & 0.840 \\
 \multicolumn{3}{l}{\textbf{Incongruent}} &&&&\\
2.5* & 0.4 & 0.4 & 0.048 & 0.140 & 0.208 & 0.260 & 0.310 \\
2.6* & 0.45 & 0.45 & 0.049 & 0.079 & 0.122 & 0.253 & 0.186 \\
2.7 & 0.2 & 0.3 & 0.720 & 0.711 & 0.544 & 0.554 & 0.453 \\
2.8 & 0.17 & 0.27 & 0.773 & 0.804 & 0.646 & 0.544 & 0.518 \\
\hline
\multicolumn{8}{c}{\textbf{Case 3: $\theta_h=0.2$}} \\
\multicolumn{3}{l}{\textbf{Congruent}} &&&&&\\
 3.1* & 0.2 & 0.2 & 0.051 & 0.050 & 0.050 & 0.050 & 0.050 \\
3.2 & 0.2 & 0.3 & 0.698 & 0.881 & 0.912 & 0.904 & 0.902 \\
3.3 & 0.21 & 0.31 & 0.696 & 0.882 & 0.922 & 0.904 & 0.926 \\
3.4 & 0.18 & 0.28 & 0.707 & 0.867 & 0.886 & 0.868 & 0.874 \\
 \multicolumn{3}{l}{\textbf{Incongruent}} &&&&\\
3.5* & 0.3 & 0.3 & 0.058 & 0.144 & 0.211 & 0.264 & 0.304 \\
3.6* & 0.35 & 0.35 & 0.054 & 0.074 & 0.136 & 0.251 & 0.190 \\
3.7 & 0.1 & 0.2 & 0.832 & 0.796 & 0.658 & 0.638 & 0.550 \\
3.8 & 0.07 & 0.17 & 0.898 & 0.876 & 0.782 & 0.635 & 0.688 \\
\hline
\hline
*Type I error.
\end{tabular}
\end{center}
\end{table}

\begin{table}[!htp]
\begin{center}
\caption {Type I error rate and power when using a non-informative prior (NP), SAM prior, mixture prior with $\tilde{w}=0.5$ (Mix50), power prior (PP), and commensurate prior (CP) with a normal endpoint. The mean of $D_h$ is $\theta_h=0$.} \label{tab:normal}
\renewcommand\arraystretch{1.12}
\begin{tabular}{ccccccccc}
\hline
\textbf{Scenario} & $\theta$ & $\theta_t$ & \textbf{NP} &  \textbf{SAM} &  \textbf{Mix50} & \textbf{PP} & \textbf{CP} \\ \hline
\multicolumn{8}{c}{\textbf{Case 1: small effect size $d=0.2$}} \\
 \multicolumn{3}{l}{\textbf{Congruent}} &&&&&\\
1.1* & 0 & 0 & 0.051 & 0.051 & 0.050 & 0.050 & 0.050 \\
1.2 & 0 & 0.6 & 0.712 & 0.910 & 0.922 & 0.912 & 0.926 \\
1.3 & -0.1 & 0.5 & 0.712 & 0.874 & 0.892 & 0.878 & 0.894 \\
1.4 & 0.1 & 0.7 & 0.712 & 0.898 & 0.932 & 0.936 & 0.949 \\
 \multicolumn{3}{l}{\textbf{Incongruent}} &&&&\\
1.5* & 0.9 & 0.9 & 0.046 & 0.070 & 0.134 & 0.326 & 0.205 \\
1.6* & 0.6 & 0.6 & 0.046 & 0.140 & 0.251 & 0.336 & 0.328 \\
1.7 & -0.6 & 0 & 0.709 & 0.652 & 0.532 & 0.526 & 0.460 \\
1.8 & -0.9 & -0.3 & 0.708 & 0.726 & 0.617 & 0.430 & 0.512 \\
\hline
\multicolumn{8}{c}{\textbf{Case 2: medium effect size $d=0.5$}} \\
\multicolumn{3}{l}{\textbf{Congruent}} &&&&&\\
2.1* & 0 & 0 & 0.051 & 0.051 & 0.050 & 0.051 & 0.051 \\
2.2 & 0 & 1.5 & 0.736 & 0.901 & 0.908 & 0.926 & 0.940 \\
2.3 & -0.2 & 1.3 & 0.734 & 0.888 & 0.892 & 0.903 & 0.916 \\
2.4 & 0.1 & 1.6 & 0.737 & 0.896 & 0.912 & 0.938 & 0.950 \\
 \multicolumn{3}{l}{\textbf{Incongruent}} &&&&\\
2.5* & 1.5 & 1.5 & 0.052 & 0.126 & 0.161 & 0.324 & 0.312 \\
2.6* & 1.8 & 1.8 & 0.052 & 0.088 & 0.139 & 0.338 & 0.264 \\
2.7 & -1.5 & 0 & 0.724 & 0.703 & 0.593 & 0.522 & 0.454 \\
2.8 & -1.8 & -0.3 & 0.722 & 0.725 & 0.606 & 0.443 & 0.433 \\
\hline
\multicolumn{8}{c}{\textbf{Case 3: large effect size $d=0.8$}} \\
\multicolumn{3}{l}{\textbf{Congruent}} &&&&&\\
3.1* & 0 & 0 & 0.051 & 0.051 & 0.050 & 0.051 & 0.050 \\
3.2 & 0 & 2.4 & 0.708 & 0.893 & 0.872 & 0.936 & 0.931 \\
3.3 & -0.3 & 2.1 & 0.704 & 0.884 & 0.860 & 0.906 & 0.912 \\
3.4 & 0.1 & 2.5 & 0.708 & 0.890 & 0.874 & 0.941 & 0.939 \\
 \multicolumn{3}{l}{\textbf{Incongruent}} &&&&\\
3.5* & 2.4 & 2.4 & 0.064 & 0.112 & 0.150 & 0.340 & 0.294 \\
3.6* & 2.7 & 2.7 & 0.066 & 0.094 & 0.136 & 0.350 & 0.262 \\
3.7 & -2.4 & 0 & 0.678 & 0.694 & 0.588 & 0.456 & 0.392 \\
3.8 & -2.7 & -0.3 & 0.672 & 0.704 & 0.592 & 0.402 & 0.411 \\
\hline
\hline
 \multicolumn{8}{l}{*Type I error; Effect size $d = (\theta_t-\theta)/\sigma$.}
\end{tabular}
\end{center}
\end{table}

\begin{table}[!htp]
\begin{center}
\caption {Type I error rate and power of the SAM prior, in comparison with the robust MAP prior with $\tilde{w}=0.5$ (Mix50) and 0.9 (Mix90), for ankylosing spondylitis trial.} \label{tab:app}
\renewcommand\arraystretch{1.12}
\begin{tabular}{ccccccccc}
\hline
\textbf{Scenario} & $\theta$ & $\theta_t$ & \textbf{NP} &  \textbf{SAM} &  \textbf{Mix50} &  \textbf{Mix90} \\ \hline
 \multicolumn{3}{l}{\textbf{Congruent}} &&&&&\\
1* & 0.36 & 0.36 & 0.050 & 0.051 & 0.050 & 0.050 \\ 
2 & 0.36 & 0.56 & 0.649 & 0.805 & 0.817 & 0.880 \\ 
3 & 0.37 & 0.57 & 0.634 & 0.821 & 0.816 & 0.897 \\ 
4 & 0.34 & 0.54 & 0.611 & 0.792 & 0.807 & 0.862 \\ 
 \multicolumn{3}{l}{\textbf{Incongruent}} &&&&\\
5* & 0.56 & 0.56 & 0.058 & 0.117 & 0.143 & 0.277 \\ 
6* & 0.61 & 0.61 & 0.053 & 0.103 & 0.128 & 0.250\\ 
7 & 0.16 & 0.36 &  0.742 & 0.679 & 0.585 & 0.463 \\ 
8 & 0.11 & 0.31 & 0.753 & 0.765 & 0.652 & 0.478 \\ 
\hline
\hline
*Type I error.
\end{tabular}
\end{center}
\end{table}

\clearpage

\begin{figure}[!htbp]
\begin{center}
\includegraphics*[scale=0.88]{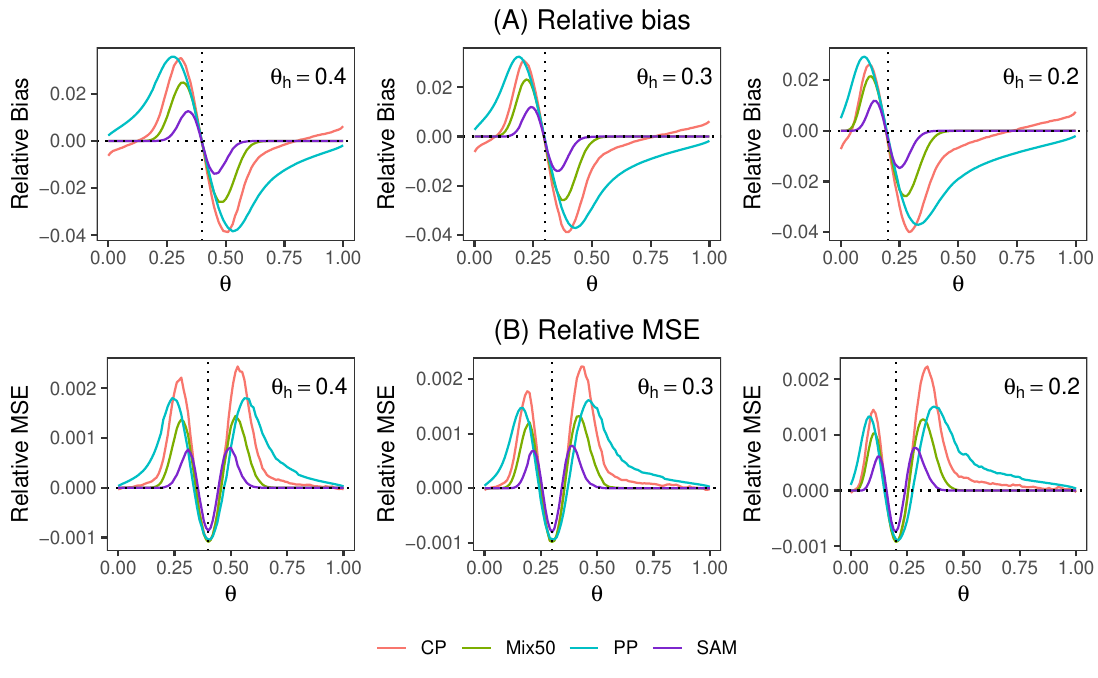}
\end{center}
\caption{(A) Relative bias and (B) relative mean square error (RMSE) for the posterior mean estimate of $\theta$ for the SAM prior, mixture prior with $\tilde{w}=0.5$ (Mix50), power prior (PP), and commensurate prior (CP) for a binary endpoint, with a non-informative prior (NP) as the reference. The vertical dotted line indicates $\theta=\theta_h$. This figure appears in color in the electronic version of this article, and any mention of color refers to that version.} \label{fig:binary}
\end{figure}

\begin{figure}[!htbp]
\begin{center}
\includegraphics*[scale=1]{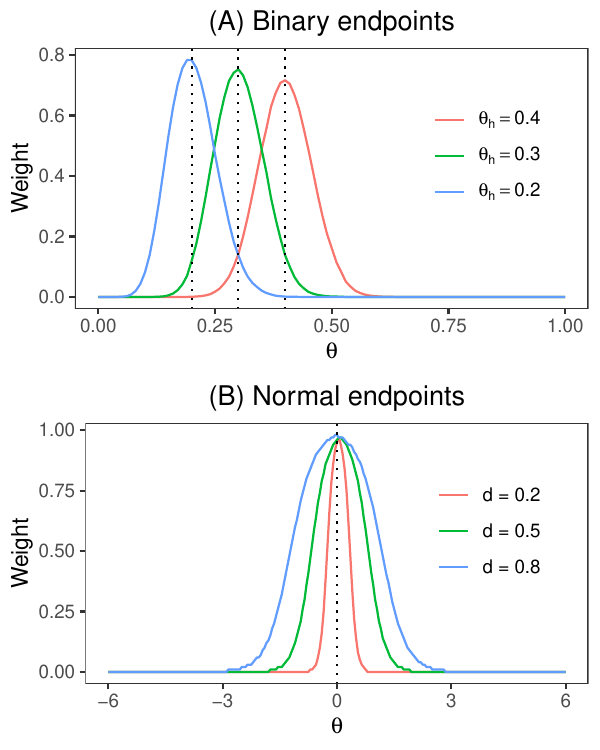}
\end{center}
\caption{Mixture weight for the SAM prior is self-adapting to prior-data conflict (i.e., $\theta-\theta_h$) for binary and normal endpoints. The vertical dotted line indicates $\theta=\theta_h$. This figure appears in color in the electronic version of this article, and any mention of color refers to that version.} \label{fig:weight}
\end{figure}

\begin{figure}[!htbp]
\begin{center}
\includegraphics*[scale=0.88]{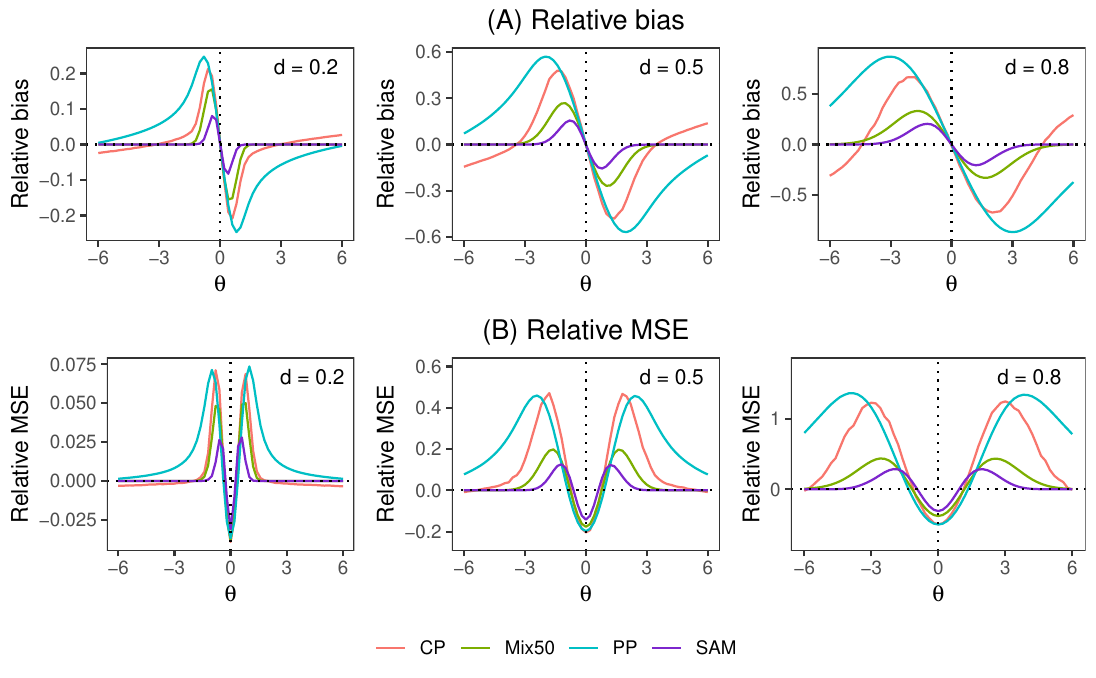}
\end{center}
\caption{(A) Relative bias and (B) relative mean square error (RMSE) for the posterior mean estimate of $\theta$ for the SAM prior, mixture prior with $\tilde{w}=0.5$ (Mix50), power prior (PP), and commensurate prior (CP) for a normal endpoint, with a non-informative prior (NP) as the reference. The vertical dotted line indicates $\theta=\theta_h$. This figure appears in color in the electronic version of this article, and any mention of color refers to that version.} \label{fig:normal}
\end{figure}

\begin{figure}[!htbp]
\begin{center}
\includegraphics*[scale=0.7]{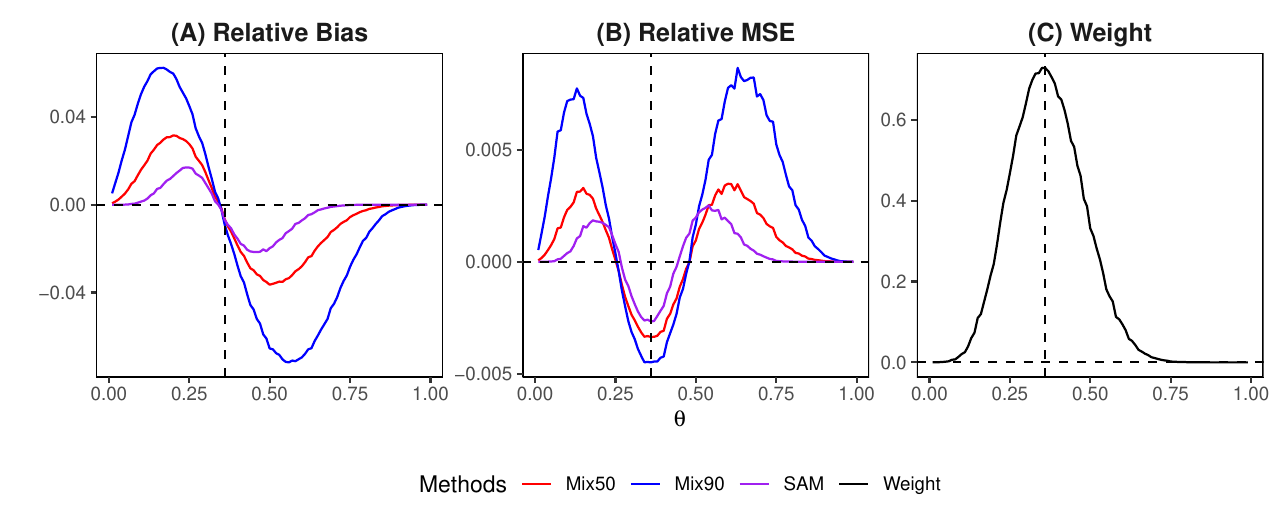}
\end{center}
\caption{ (A) Relative bias and (B) relative mean square error (RMSE) for the posterior mean estimate of $\theta$ for the SAM prior and robust MAP prior with $\tilde{w}=0.5$ (Mix50) and 0.9 (Mix90) in ankylosing spondylitis trial. Panel (C) depicts how the weight of the SAM prior self-adapts to prior-data conflict (i.e., $\theta-\theta_h$). The vertical dotted line indicates $\theta=\theta_h$. This figure appears in color in the electronic version of this article, and any mention of color refers to that version.} \label{fig:app}
\end{figure}

\backmatter

\label{lastpage}

\end{document}